\input{aipcheck}

\documentclass[
    ,final            
  ]
  {aipproc}

\layoutstyle{8x11single}

\begin{document}

\title{Bound $H$-dibaryon in the Flavor $SU(3)$ Limit \\ from a Full QCD Simulation on the Lattice}



\author{Takashi Inoue (for HAL QCD Collaboration)}{
  address={Nihon University, College of Bioresource Sciences, Kanagawa 1866, Japan}
}

\begin{abstract}
 Existence of the $H$-dibaryon in the flavor $SU(3)$ symmetric limit is studied by full QCD simulations on the lattice,
 in the approach recently developed for the baryon-baryon ($BB$) interactions.
 Potential of the flavor-singlet $BB$ channel is derived from the Nambu-Bethe-Salpeter wave function,
 and a bound $H$-dibaryon is discovered from it,
 with the binding energy of 20--50 MeV for the pseudo-scalar meson mass of 469--1171 MeV.


\end{abstract}

\maketitle


\section{Introduction}

 The $H$-dibaryon, predicted by R.~L.~Jaffe in 34 years ago~\cite{Jaffe:1976yi}.
 is one of most famous candidates of exotic-hadron. 
 This prediction was based on the observations that the quark exclusion can
 be completely avoided due to the essentially flavor-singlet($uuddss$)
 nature of $H$-dibaryon, together with the large attraction from
 one-gluon-exchange interaction between quarks suggested in the quark
 model~\cite{Jaffe:1976yi,Sakai:1999qm}.
 Search for the $H$-dibaryon is one of the most challenging theoretical and experimental problems
 in the physics of strong interaction and quantum chromodynamics (QCD).
 Still it is not clear whether there exists the $H$-dibaryon in nature.
 Although deeply bound $H$-dibaryon with the binding energy $B_H > 7 $ MeV from the $\Lambda\Lambda$ threshold
 has been ruled out by the discovery of the double $\Lambda$ hypernuclei,
 $_{\Lambda \Lambda}^{\ \ 6}$He~\cite{Takahashi:2001nm}, there still  remains a possibility of 
 a shallow bound state or a resonance in this channel~\cite{Yoon:2007aq}.
 
 Several lattice QCD calculations on $H$ have been reported as reviewed in~\cite{Wetzorke:2002mx}
 (see also recent works \cite{Luo:2007zzb,Inoue:2010hs,Beane:2010hg}).
 However, there is a serious problem in studying dibaryons on the lattice:
 To accommodate two baryons inside the lattice volume,
 the spatial lattice size $L$ should be large enough. 
 Once  $L$ becomes large, however,  energy levels of two baryons become dense,
 so that isolation of the ground state from the excited states is very difficult
 (quite a large imaginary-time $t$ is required).
 All the previous works on dibaryons more or less face this issue.
 In this paper, we employ our original approach distinct from one in previous studies,
 and search the $H$-dibaryon in lattice QCD ~\cite{Inoue:2010es}.
 Because the flavor $SU(3)$ breaking complicate calculation, we start with the flavor $SU(3)$ limit.

\section{Formalism}
 To study interaction or property of a hadron systems in lattice QCD,
 we utilize the Nambu-Bethe-Salpeter (NBS) wave function and the corresponding potential,
 instead of the energy eigenstate.
 Within the non-relativistic approximation, 
 potential of a two-body interaction can be defined from the NBS wave function $\Psi(\vec r, t)$ 
 through the Schr\"{o}dinger equation in the Euclidean space-time \cite{Ishii:2006ec,Nemura:2008sp,Aoki:2009ji},
\begin{equation}
      \left[ M_1 + M_2 - \frac{\nabla^2}{2\mu} \right] \Psi(\vec r, t)
    + \int \!\! d^3 \vec r' \, U(\vec r, \vec r') \, \Psi(\vec r', t) 
    = - \frac{\partial}{\partial t} \Psi(\vec r, t)
\end{equation}
where a non-local but energy-independent potential $U(\vec r, \vec r')$ is introduced,
and $\mu$ is the reduced mass of the system with masses $M_1$ and $M_2$.
The potential $U(\vec r, \vec r')$ can be determined by solving this equation 
by using data of $\Psi(\vec r, t)$ measured on the lattice.
Since only a single data of $\Psi(\vec r, t)$ is available from one simulation,
we cannot construct whole the potential $U(\vec r, \vec r')$,
but can obtain the leading term of the velocity expansion of it. 
We describe the potential $V(r)$. 
As shown in ref. \cite{Inoue:2010es}, one obtains a sink-time independent $V(r)$ even without isolating energy eigenstates.
It is also shown that we can obtain a volume independent $V(r)$,
if we setup the volume larger than the minimum to accommodate the interaction,
e.g. $L \simeq 4$ fm for the nucleon-nucleon interaction at $M_{\pi} \ge 450$ MeV.
Once we obtain a volume independent potential,
any observable of the system, such as binding energy and scattering phase shift,
can be obtained by solving the Schr\"{o}dinger equation in the infinite volume.
Note that, in contrast to the conventional L\"{u}scher's method \cite{Luscher:1990ux,Fukugita:1994ve},
we do not calculate the energy shift of two hadrons at finite $L$
to access the observables in the infinite volume.
Finally, the NBS wave function can be obtained by summing up the hadron four-pint function $G_4$ 
which can be evaluated numerically in lattice QCD.

\section{Lattice QCD setup}

\begin{table}[t]
\caption{\label{tbl:lattice}Summary of lattice parameters and hadron masses.}
 \begin{tabular}{c|c|c|c|c|c|c}
   \hline
  $a$ [fm] & $L$ [fm] &  $\kappa_{uds}$  & ~$M_{\rm p.s.}$ [MeV]~ &  ~$M_{\rm vec}$ [MeV]~& ~ $M_{\rm bar}$ [MeV]~ & ~$N_{\rm cfg}$~ \\
   \hline 
              &      &   ~0.13660~ &   1170.9(7) &   1510.4(0.9) & 2274(2) & 420 \\
              &      &   ~0.13710~ &   1015.2(6) &   1360.6(1.1) & 2031(2) & 360 \\
   {0.121(2)} & 3.87 &   ~0.13760~ & ~\,836.5(5) &   1188.9(0.9) & 1749(1) & 480 \\
              &      &   ~0.13800~ & ~\,672.3(6) &   1027.6(1.0) & 1484(2) & 360 \\
              &      &   ~0.13840~ & ~\,468.9(8) & ~\,830.6(1.5) & 1163(2) & 600 \\

   \hline
 \end{tabular}
\end{table}

 For dynamical lattice QCD simulations, we have generated five ensembles of gauge configuration on $32^3 \times 32$ lattice. 
 We have employed the renormalization group improved Iwasaki gauge action with the coupling constant $\beta=1.83$, 
 and the non-perturbatively $O(a)$ improved Wilson quark action.  
 Quark propagators are calculated for the quark wall source
 with the Dirichlet boundary condition in the temporal direction.
 The point type octet-baryon field operators are used at sink.
 The sink $BB$ operator is projected to the $A_{1}^{+}$ representation of the cubic group,
 so that the NBS wave function is dominated by the S-wave component.
 For the time derivative of the NBS wave function, we adopt the symmetric difference on the lattice. 
 To enhance signal over noise, we average 16 measurements for each configuration,
 together with the average between forward and backward propagation in time.
 Statistical errors are evaluated by the Jackknife method.
 Lattice parameters such as lattice spacing $a$, lattice size $L$, the hopping parameter $\kappa_{uds}$,
 the number of gauge configurations $N_{\rm cfg}$,  together with the hadron masses are summarized in Table \ref{tbl:lattice}.

\section{Results}

\begin{figure}[t]
\includegraphics[width=8.0cm]{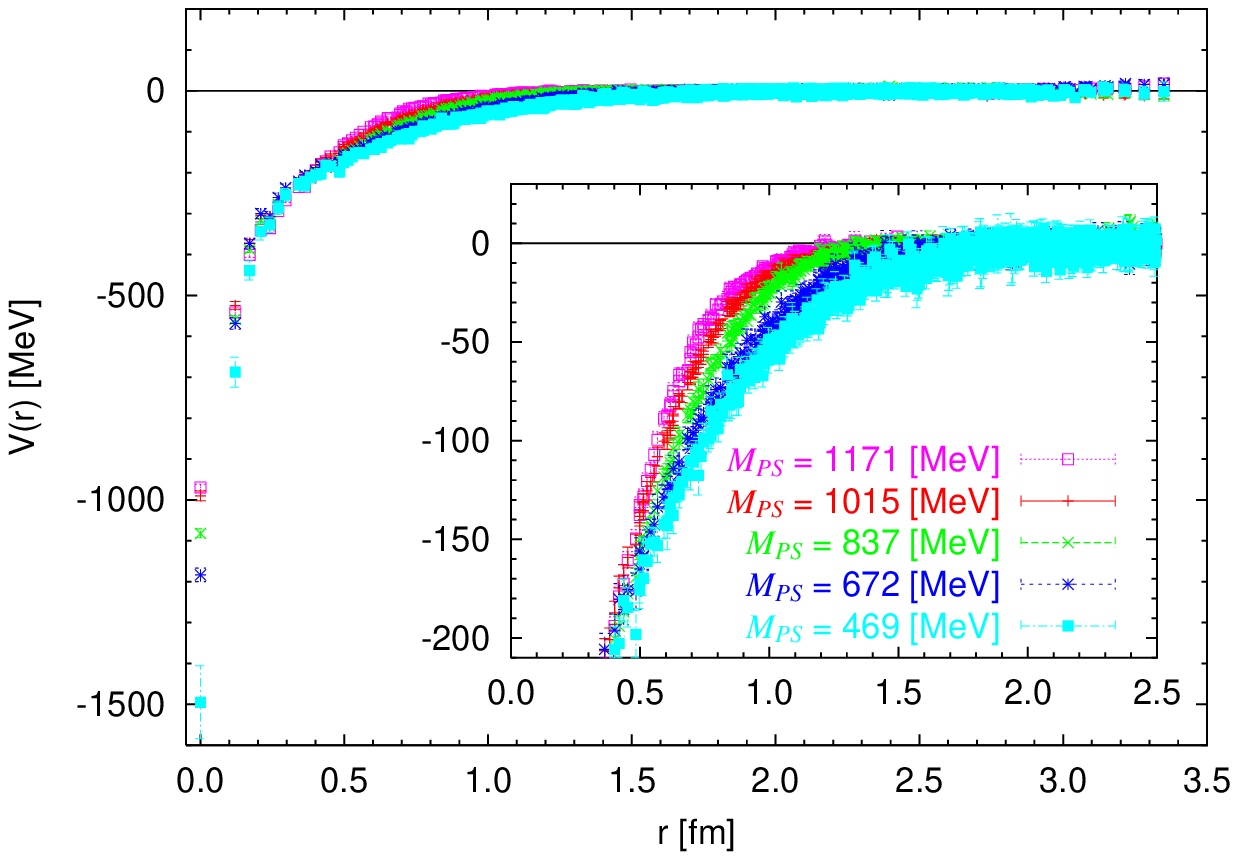} \quad
\includegraphics[width=8.0cm]{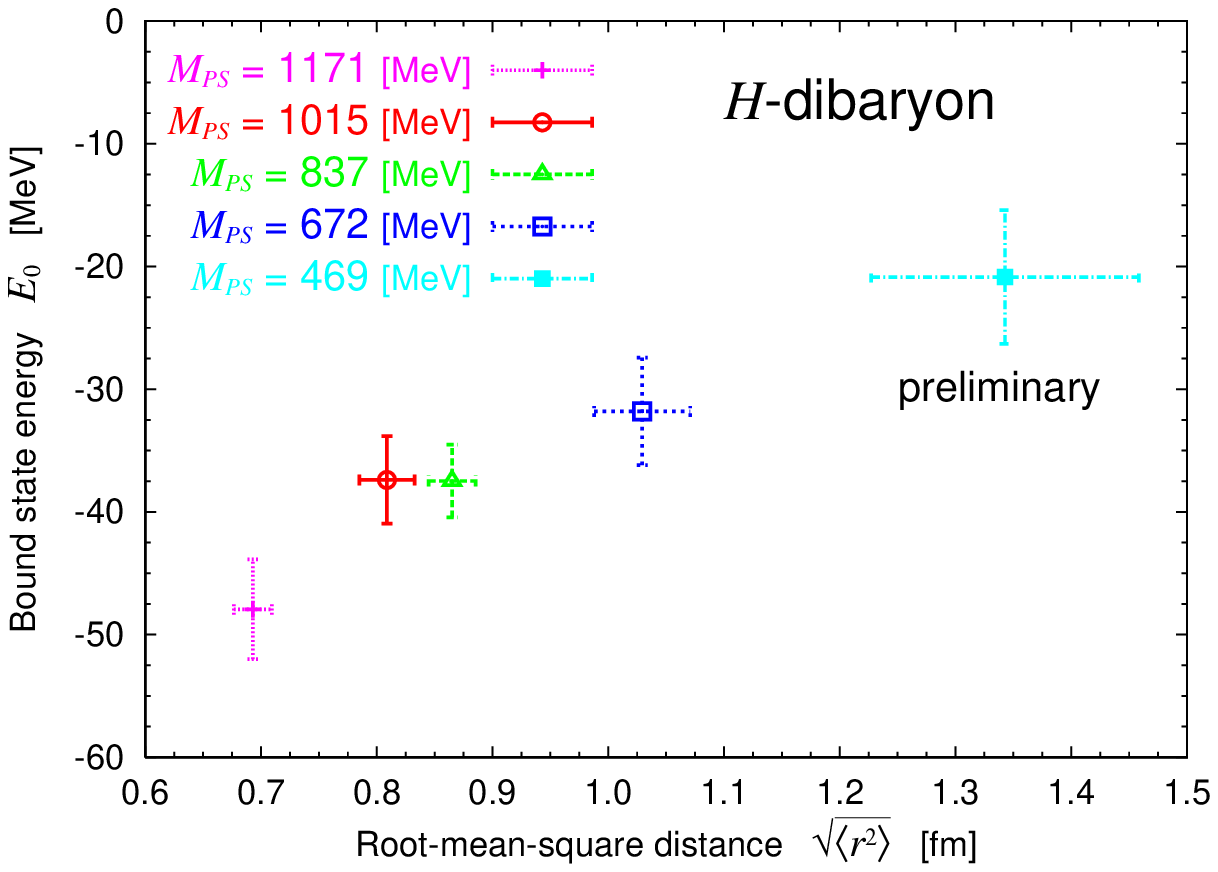}
\caption{ Left:  Potential of the flavor-singlet $BB$ channel.
          Right: The ground state of the flavor-singlet $BB$ channel.}
\label{Fig1}
\end{figure}

 The left panel of Fig.\ref{Fig1} shows the flavor-singlet $BB$ potential at each quark mass,
 extracted from our numerical simulation of full QCD.
 One see that it has an ``attractive core" and its range is well localized in space.
 This entirely attractive potential prove that the quark model prediction to the flavor-singlet $BB$ channel
 is essentially correct.
 One can see that the long range part of the attraction tends to increase as the quark mass decreases.

 By solving the Schr\"{o}dinger equation involving this potential, 
 we can obtain any information of the system, for instance, energy of the ground state.
 It turns out that there is only one bound state in each quark mass~\cite{Inoue:2010es}.
 The right panel of Fig.\ref{Fig1} shows the energy and the root-mean-square (rms) distance of the bound state.
 The result for $M_{\rm p.s.}=469$ MeV is only preliminary because of insufficient number of independent gauge configuration.
 These bound state correspond to the $H$-dibaryon predicted by Jaffe.
 Our lattice QCD calculation shows that a stable $H$-dibaryon certainly exists
 in the flavor $SU(3)$ limit world with the binding energy of 20--50 MeV for the present quark masses. 
 Despite that the attractive potential become stronger as quark mass decreases,
 the resultant binding energies of the $H$-dibaryon decrees in the present range of the quark masses.
 This is due to the fact that the increase of the attraction toward the lighter quark mass
 is compensated by the increase of the kinetic energy for the lighter baryon mass. 
 The rms distance $\sqrt{\langle r^2 \rangle}$ is a measure of spacial distribution of baryonic matter in the $H$-dibaryon,
 which can be compared to the point matter rms distance of the deuteron in nature, $1.9 \times 2 = 3.8$ fm.
 One may be able to get feeling of the $H$-dibaryon from this comparison:
 $H$-dibaryon is much compact compared to usual nuclei.

 When we solve the Schr\"{o}dinger equation, 
 we use the potential expressed in terms of an analytic function fitted to the data. 
 Error from the choice for the analytic function to fit is negligible (less than few \%).  
 Small systematic error arise from choice for the sink-time-slice.
 The final result of the binding energy of the $H$-dibaryon in the $SU(3)$ limit
 described as ${\tilde B}_H$ becomes, for example,
\begin{equation}
 M_{\rm p.s.}=~837 ~\mbox{MeV} :~ \  {\tilde B}_H = 37.8 (3.1)(4.2) ~\mbox{MeV} 
\end{equation} 
 with statistical error (first) and systematic error (second).

One may think such a deeply bound $H$-dibaryon is ridiculous or ruled out by the discovery of 
the double $\Lambda$ nuclei.
The present $\tilde{B}_H$ should be interpreted as the binding energy from the $BB$ threshold averaged in the $(S,I)=(-2,0)$ sector.
The three $BB$ thresholds in this sector split largely in the real world with the flavor $SU(3)$ breaking. 
Therefore, we expect that the binding energy of the $H$-dibaryon in the real world, measured from the $\Lambda\Lambda$ threshold,
is much smaller than the present value or even $H$-dibaryon goes to above the $\Lambda\Lambda$ threshold.
To make a definite conclusion on this point, however, we need (2+1)-flavor lattice QCD simulations
and the analysis in $\Lambda\Lambda-N\Xi-\Sigma\Sigma$ coupled channel. 
Study along this direction is in progress~\cite{sasaki2010}. 

\medskip


\begin{theacknowledgments}
We thank K.-I. Ishikawa and PACS-CS group for providing their DDHMC/PHMC code~\cite{Aoki:2008sm},
and authors and maintainer of CPS++~\cite{CPS} for a modified version of it used in this paper.
Numerical computations of this work have been carried out at Univ. of Tsukuba supercomputer system (T2K).
This research is supported in part by MEXT Grant-in-Aid for Scientific Research on Innovative Areas(No.2004:20105001, 20105003).
\end{theacknowledgments}




\bibliographystyle{aipproc}   

\begin{thebibliography}{99}


\bibitem{Jaffe:1976yi}
  R.~L.~Jaffe,  
  Phys.\ Rev.\ Lett.\  {\bf 38}, 195 (1977)
  [Erratum-ibid.\  {\bf 38}, 617 (1977)].

\bibitem{Sakai:1999qm}
  T.~Sakai, K.~Shimizu and K.~Yazaki,
  Prog.\ Theor.\ Phys.\ Suppl.\  {\bf 137}, 121 (2000)
  [arXiv:nucl-th/9912063].

\bibitem{Takahashi:2001nm}
  H.~Takahashi {\it et al.},
  Phys.\ Rev.\ Lett.\  {\bf 87}, 212502 (2001).
  
 \bibitem{Yoon:2007aq}
  C.~J.~Yoon {\it et al.},
  Phys.\ Rev.\  C {\bf 75}, 022201 (2007).


\bibitem{Wetzorke:2002mx}
  I.~Wetzorke and F.~Karsch,
  Nucl.\ Phys.\ Proc.\ Suppl.\  {\bf 119}, 278 (2003)
  [arXiv:hep-lat/0208029].

\bibitem{Luo:2007zzb}
  Z.~H.~Luo, M.~Loan and X.~Q.~Luo,
  Mod.\ Phys.\ Lett.\  A {\bf 22}, 591 (2007)
  [arXiv:0803.3171 [hep-lat]].

\bibitem{Inoue:2010hs}
  T.~Inoue {\it et al.}  [HAL QCD Coll.],
  Prog. Theor. Phys. {\bf 124}, 591 (2010)
  [arXiv:1007.3559 [hep-lat]].


\bibitem{Beane:2010hg}
  S.~R.~Beane {\it et al.}  [NPLQCD Collaboration],
  Phys.\ Rev.\ Lett.\  {\bf 106}, 162001 (2011)
  [arXiv:1012.3812 [hep-lat]].

\bibitem{Inoue:2010es}
  T.~Inoue {\it et al.}  [HAL QCD Collaboration],
  Phys.\ Rev.\ Lett.\  {\bf 106}, 162002 (2011)
  [arXiv:1012.5928 [hep-lat]].

\bibitem{Ishii:2006ec}
  N.~Ishii, S.~Aoki and T.~Hatsuda,
  Phys.\ Rev.\ Lett.\  {\bf 99}, 022001 (2007)
  [arXiv:nucl-th/0611096].
\bibitem{Nemura:2008sp}
  H.~Nemura, N.~Ishii, S.~Aoki and T.~Hatsuda,
  Phys.\ Lett.\  B {\bf 673}, 136 (2009)
  [arXiv:0806.1094 [nucl-th]].
\bibitem{Aoki:2009ji}
  S.~Aoki, T.~Hatsuda and N.~Ishii,
  Prog.\ Theor.\ Phys.\  {\bf 123}, 89 (2010)
  [arXiv:0909.5585 [hep-lat]].
 



\bibitem{Luscher:1990ux}
  M.~L\"{u}scher,
  Nucl.\ Phys.\  B {\bf 354}, 531 (1991).

\bibitem{Fukugita:1994ve}
  M.~Fukugita {\it et al.},
  Phys.\ Rev.\  D {\bf 52}, 3003 (1995)
  [arXiv:hep-lat/9501024].

 \bibitem{sasaki2010} 
 K.~Sasaki [HAL QCD Coll.], PoS {\bf LAT 2010}, 157 (2010).

\bibitem{Aoki:2008sm}
  S.~Aoki {\it et al.}  [PACS-CS Coll.],
  Phys.\ Rev.\  D {\bf 79}, 034503 (2009)
  [arXiv:0807.1661 [hep-lat]].

\bibitem{CPS}
Columbia Physics System (CPS), http://qcdoc.phys.columbia.edu/cps.html

\end{thebibliography}


\end{document}